\documentclass[twocolumn,showpacs,preprintnumbers,amsmath,amssymb]{revtex4}

\usepackage{graphicx}
\usepackage{dcolumn}
\usepackage{bm}
\usepackage{latexsym}

\begin{document}

\preprint{APS/paper}

\title{Concentration phase diagram of Ba$_{x}$Sr$_{1-x}$TiO$_{3}$ solid solutions}

\author{V.B. Shirokov}
\author{V.I. Torgashev}
\author{A.A. Bakirov}

\affiliation{%
Research Institute of Physics, Rostov State University, 344090 Rostov on Don, Russia}%
\author{V. V. Lemanov}
\affiliation{A. F. Ioffe Physico-Technical Institute, 194021 St. Petersburg, Russia}%
\date{\today}

\begin{abstract}
Method of derivation of phenomenological thermodynamic potential of solid solutions is proposed in which the interaction of the order parameters of constituents is introduced through the account of elastic strain due to misfit of the lattice parameters of the end-members. The validity of the method is demonstrated for Ba$_{x}$Sr$_{1-x}$TiO$_{3}$ system being a typical example of ferroelectric solid solution. Its phase diagram is determined using experimental data for the coefficients in the phenomenological potentials of SrTiO$_3$ and BaTiO$_3$. In the phase diagram of the Ba$_{x}$Sr$_{1-x}$TiO$_{3}$ system for small Ba concentration, there are a tricritical point and two multiphase points one of which is associated with up to 6 possible phases.
\end{abstract}
\pacs{61.50.Ks, 64.60.Kw, 64.70.Kb, 77.80.Bh, 77.84.Dy }
\maketitle

\section{\label{sec:level1}INTRODUCTION}
The solid solution of BaTiO$_{3}$ with SrTiO$_{3}$ is one of the best documented systems of ferroelectric solid solutions (see Refs. \cite{c1,c2,c3} and references therein).

BaTiO$_{3}$ (BT in short) is well known as the first oxygen-octahedra ferroelectric of the perovskite structure with paraelectric cubic phase Pm3m and three ferroelectric phases: tetragonal P4mm, orthorhombic Amm2, and rhombohedral R3m. All three transitions are described by one three-component order parameter (polarization), and with the potential expansion up to the sixth order \cite{c5}. Best agreement with the experiment is achieved when not only one constant in the term quadratic in the order parameter is assumed to be temperature-dependent \cite{c7}.

SrTiO$_{3}$  (ST in short) represents a textbook example of a "quantum paraelectric" \cite{c8}  with a polar soft mode, which has never condensed even at very low temperatures. At a temperature T$_a$  of about 105 K pure ST undergoes an antiferrodistortive (improper ferroelastic) phase transition Pm3m $\to$ I4/mcm with doubling of the primitive cell volume. Microscopic mechanism of the transition is attributed to the instability of a soft lattice mode associated with out-of-phase rotations of TiO$_{6}$ octahedra. The transition is phenomenologically described by three-component order parameter transforming according to $R_{25}$ representation from the point R ( 1/2, 1/2, 1/2) of the Brillouin zone (BZ) of a simple cubic lattice with the Landau potential of the fourth order \cite{c9}. The constants of the potential were found in Refs. \cite{c10,c11,c12}.

Along with softening the soft mode from the BZ boundary on cooling \cite{c13,c14}, the static dielectric permittivity significantly increases \cite{c8,c15} due to softening of IR-active $F_{1u}$ mode from the BZ center \cite{c16,c17}. Down to 1.5 K, the ST crystals remain in the tetragonal phase \cite{c8,c18}. To describe the temperature behavior of the static dielectric permittivity, the Barrett quantum equation \cite{c19} is applied \cite{c8}. Thus, the phenomenological potential, which simultaneously describes the above-mentioned properties should include two three-component order parameters $R_{25}  \oplus F_{1u} $. The constants of the fourth order potential with such order parameters were detemined in Refs.  \cite{c20,c21,c22}.

According to a Curie-Weiss law in ST, wich follows from dielectric and optical measurements \cite{c8,c16,c17,c23,c26}, a  ferroelectric phase transition should occur at a temperature near 35 K  associated with softening of the IR-active mode. No dielectric anomaly at this temperature was found but many authors observed elastic anomalies \cite{c27,c28,c29,c30,c31,c32,c33,c34}. However, in careful dilatometric studies, no low-temperature anomalies were observed \cite{c35}.

The solid solution Ba$_{x}$Sr$_{1-x}$TiO$_{3}$ (BST in short) of these two remarkable compounds attracts attention of researches from at least two aspects: 1) basic interest concerning such problems as impurity induced ferroelectric phase transition in the ST incipient ferroelectric, as interplay between the antiferrodistortive and ferroelectric phase transitions, as development of the theory to predict the  $(x - T)$  phase diagram; 2) applied aspect as the possibility to control properties and the phase transition temperature of these promising for various application materials.

At first, the Ba$_{x}$Sr$_{1-x}$TiO$_{3}$ solid solutions have been studied at $x > 0.1$ and at temperatures more than about 100 K. Hegenbarth \cite{c36}  and Bednorz \cite{c37} studied imputity-induced ferroelectric phase transitions in the BST solid solutions at small $x$ and down to liquid helium  temperatures. Bednorz \cite{c37} and Miura et al. \cite{c38} investigated the effect of Ba substitution on the antiferrodistortive  phase transition. Later on, this system at low $x$ was studied in more details \cite{c39,c40,c41,c42}.

The summary of all the experimental data concerning the $(x-T)$ phase diagram of BST is the following.

The temperatures of all three ferroelectric phase transition decrease with $x$ decreasing with different rates and probably converge at $x$ around 0.1.

At $x = 0.2\div 0.25$ \cite{c39} or $x = 0.4$ \cite{c43}, there is a tricritical point for the cubic - tetragonal phase transition when the first order phase transition with $x$ decreasing transforms to the second order one.

The antiferrodistortive transition temperature T$_a$ decreases with $x$ increasing, and this transition disappears when T$_a$ should become lower than ferroelectric transition temperature \cite{c42}.

At small $x$ ( $x = 0.005$ \cite{c37, c40} and $x = 0.01\div0.03$ \cite{c36, c39, c42} ), the BST system becomes ferroelectric, and  the transition temperature  T$_c$ increases with $x$ increasing.

In spite of this rich experimental data on the BST $(x - T)$ phase diagram, no theoretical treatment of the phase diagram is available. This problem is, as mentioned above, of great interest both basic and applied, and in this work we develop the phenomenological theory and derive the concentration phase diagram for BST solid solutions. The region of small concentration of Ba (small $x$) is especially interesting since just in the region of small concentration: i) the impurity-induced ferroelectricity occurs; ii) there is a strong interplay between ferroelastic (antiferrodistortive) and ferroelectric phase transitions; iii) experimental data on the symmetry of low-temperature phases are not available.

The paper is organized as follows. After Introduction, where background of the problem is depicted, Sec. II follows with derivation of the phenomenological potential of the BST solid solution. In Sec. III , the $(x-T)$ phase diagram of the BST system is analyzed, and specific features of the phase diagram are discussed in Sec. IV. The last Section presents a summary of the work.

\section{ PHENOMENOLOGICAL POTENTIAL OF THE $\bm{\mathbf{Ba_{x}Sr_{1-x}TiO_3}}$ SOLID SOLUTION }

\subsection{ Formalism }

Thermodynamic description of solid solutions we shall make by the expansion of the Helmholtz free energy in powers of order parameter of  a high-symmetry phase \cite{c4}. At all temperatures, we treat the system as a homogeneous solid solution, which does not exhibit the ordering of its components. The constants of the free energy expansion are temperature- and concentration-dependent. We suppose that we know the potentials of the solid soluton end-members with concentration $x = 0$ and $x = 1$. Moreover, we suppose that at least at high temperatures the end-members exist in one and the same polymorphic modification. We shall derive the thermodynamic potential of the solid solution of arbitrary concentration using the following approaches.

Atom substitution with no phase transition leads to a homogeneous change of crystal size. We shall treat the crystal of solid solution as a mixture of two interacting crystals with concentration $x = 0$ and $x = 1$ with
molar contribution $(1 - x)$ and $x$, respectively.  The interaction of such 'crystals' will be described not by usual introduction of additive terms in the thermodynamic potential but through the account of elastic strains remaining in the framework of the phenomenological theory. These strains for the solid solution end-members  $u_0$ and  $u_1$   will be determined using two conditions. The first one is the fit of the lattice parameter of the solid solution  $a_x$ and that of "deformed" end-members ($x$ = 0 and $x$ = 1).

Let us express the (inequilibrium) lattice parameter of the solid solution $a$ via its equilibrium value $a_x$ and the volume strain $u$:

\[a = a_x (1 + u)\]

Then, the lattice parameters of the solid solution end members are

\begin{equation}
\begin{array}{l}
 a_0  = a_x (1 + u_0 ), \\
 a_1  = a_x (1 + u_1 ). \\
 \end{array}
\end{equation}

The elastic energy of the crystals $x = 0$ and $x = 1$ are $F_0$ and  $F_1$ respectively. The strain in the $F_0$ potential may be expressed via the strain $u$ of the solid solution

\[
\frac{{a - a_0}}
{{a_0}} = \frac{{a_x (1 + u) - a_x (1 + u_0 )}}
{{a_x (1 + u_0 )}} = \frac{{u - u_0 }}
{{1 + u_0 }} \approx u - u_0
\]

Similar relation may be written for  the $F_1$ potential. As a result, the part of the elastic energy related to the crystals $x = 0$ and $x = 1$ will be $(1-x)F_0(u-u_0)$  and $xF_1(u-u_1)$  respectively. Since the interaction is introduced through deformation, the elastic energy of the solid solution can be written as

\[(1-x)F_0(u-u_0) + xF_1(u-u_1)\]

Consequently, the second condition is the condition of equilibrium obtained by minimization 
of this energy with respect to the strain $u$. As $a=a_x$ in the equilibrium ($u = 0$) we get the equation

\begin{equation}
\left. {(1 - x)\frac{{\partial F_0 }}{{\partial u}}} \right|_{u = 0 }  + \left. {x\frac{{\partial F_1 }}{{\partial u}}} \right|_{u = 0 }  = 0
\end{equation}

It means that the resulting stress should be zero when lattice parameter of the solid solution acquires its equilibrium value $a_x$. The equations (1), (2) allow to find the values of deformations $u_0$ and $u_1$, as well as the equilibrium value of the solid solution lattice parameter $a_x$.

Writing the $F_0$ and $F_1$ potentials in the solid solution lattice by means of deformanion, one can introduce, along with the common strain  $u$, and common order parameters $\eta$. As a result, the potential of the solid solution written through the potentials of its end-members will be given as

\begin{equation}
F = (1 - x)F_0 (\eta ,u - u_0 ) + xF_1 (\eta ,u - u_1 )
\end{equation}

To proceed further, one should minimize Eq.(3) with the account of Eqs.(1), (2).

\subsection{Potential of the BST system}

We apply the approach developed in the previous subsection to the Ba$_{x}$Sr$_{1-x}$TiO$_{3}$ solid solution (now $x = 0$ and $x = 1$  mean ST and BT, respectively). The Helmholtz free energy per unit  volume $F_i$ (i= ST or BT) is given as \cite{c20}

\begin{widetext}
\begin{equation}
\begin{array}{l}
 F_i  = \beta _{1,i}^u \left( {\varphi _1^2  + \varphi _2^2  + \varphi _3^2 } \right) + \beta _{11,i}^u \left( {\varphi _1^4  + \varphi _2^4  + \varphi _3^4 } \right) + \beta _{12,i}^u \left( {\varphi _1^2 \varphi _2^2  + \varphi _1^2 \varphi _3^2  + \varphi _2^2 \varphi _3^2 } \right) + \alpha _{1,i}^u \left( {p_1^2  + p_2^2  + p_3^2 } \right)\\[6pt]
 \qquad  + \alpha _{11,i}^u \left( {p_1^4  + p_2^4  + p_3^4 } \right) + \alpha _{12,i}^u \left( {p_1^2 p_2^2  + p_1^2 p_3^2  + p_2^2 p_3^2 } \right) + \alpha _{111,i}^u \left( {p_1^6  + p_2^6  + p_3^6 } \right) + \alpha _{112,i}^u [p_1^4 (p_2^2  + p_3^2 )\\[6pt]
 \qquad   + p_2^4 (p_1^2  + p_3^2 ) + p_3^4 (p_1^2  + p_2^2 )] + \alpha _{123,i}^u p_1^2 p_2^2 p_3^2  - t_{11,i}^u \left( {\varphi _1^2 p_1^2  + \varphi _2^2 p_2^2  + \varphi _3^2 p_3^2 } \right) - t_{12,i}^u [\varphi _1^2 (p_2^2  + p_3^2 )\\[6pt]
 \qquad    + \varphi _2^2 (p_1^2  + p_3^2 ) + \varphi _3^2 (p_1^2  + p_2^2 )] - t_{44,i}^u \left( {\varphi _2 \varphi _3 p_2 p_3  + \varphi _1 \varphi _3 p_1 p_3  + \varphi _1 \varphi _2 p_1 p_2 } \right)  \quad + \quad F_{u,i}.  \\[6pt]
\end{array}
\end{equation}
\end{widetext}

Here, $\varphi $ is the rotational order parameter (oxygen ion shift in the $R_{25}$ mode related to out-of phase rotation of TiO$_{6}$ octahedral); $p$ , the polarization (related to polar shifts of ions in the $F_{1u}$ mode); $F_{u,i}$ , the deformation potential presented below. The superscript $u$ in Eq. (4) indicates that the coefficients are taken at constant strain. We shall further suppose that the elastic strains are not the order parameters, i.e., there are no proper ferroelastic transitions. Then, the deformation potential should be quadratic in strain

\begin{widetext}
\begin{equation}
\begin{array}{l}
 F_{u,i}  =  - (c_{11,i}  + 2c_{12,i} )\alpha _i T(u_1  + u_2  + u_3 ) + c_{12,i} (u_1 u_2  + u_1 u_3  + u_2 u_3 ) +  \frac{1}{2}c_{11,i} (u_1^2  + u_2^2  + u_3^2 )\\[6pt]
  \qquad  + \frac{1}{2}c_{44,i} (u_4^2  + u_5^2  + u_6^2 ) - b_{11,i} (u_1 \varphi _1^2  + u_2 \varphi _2^2  + u_3 \varphi _3^2 ) - b_{44,i} (u_4 \varphi _2 \varphi _3  + u_5 \varphi _1 \varphi _3  + u_6 \varphi _1 \varphi _2 )\\[6pt]
  \qquad - b_{12,i} [u_1 (\varphi _2^2  + \varphi _3^2 ) + u_2 (\varphi _1^2  + \varphi _3^2 ) + u_3 (\varphi _1^2  + \varphi _2^2 )]  - g_{11,i} (u_1 p_1^2  + u_2 p_2^2  + u_3 p_3^2 ) \\[6pt]
  \qquad - g_{44,i} (u_4 p_2 p_3  + u_5 p_1 p_3  + u_6 p_1 p_2 ) - g_{12,i} [u_1 (p_2^2  + p_3^2 ) + u_2 (p_1^2  + p_3^2 ) + u_3 (p_1^2  + p_2^2 )], \\[6pt]
\end{array}
\end{equation}
\end{widetext}

where $c_{kj,i}$ are the elastic moduli of the i-th components; $\alpha_i$, their thermal expansion coefficients (i = ST, BT); T, the absolute temperature, $u$ is the strain: $u_k  = \frac{{\partial \left( {\delta x_k } \right)}}{{\partial x_k }}$, k=1, 2, 3; $u_4  = \frac{{\partial \left( {\delta x_2 } \right)}}{{\partial x_3 }} + \frac{{\partial \left( {\delta x_3 } \right)}}{{\partial x_2 }}$, \ldots , in the Voigt notation . Only linear in strain and quadratic in the $\varphi$  and $p$ interaction terms are included in Eq. (5) \cite{c20}.

Now, we write the phenomenological potential of the solid solution. From Eq. (3), it follows

\begin{equation}
\begin{array}{l}
 F_{BST}  = (1 - x)F_{ST} (\varphi ,p,u_k  - \Delta _{ST} ,u_m ) \\[6pt]
 \qquad \qquad + \quad  xF_{BT} (\varphi ,p,u_k  - \Delta _{BT} ,u_m ) \\[6pt]
\end{array}
\end{equation}
where  $k$ = 1, 2, 3,  $m$ = 4, 5, 6, and $\Delta_{ST}$, $\Delta_{BT}$ are determined by Eqs. (1), (2) where $u_0=\Delta_{ST}$, $u_1=\Delta_{BT}$ with the deformation potentials Eq.(5) at $\varphi=0$, $p=0$. In the high-symmetry phase Eqs. (1), (2) yield

\begin{equation}
\begin{array}{l}
 a_x = \frac{{(1 - x)\tau a_{ST}  + xa_{BT} }}{{(1 - x)\tau  + x - [ {(1 - x)\tau  + x(1 + \gamma )} ] \alpha _{ST} T}} \\[6pt]
 \Delta _{ST}  = \frac{{ - x\delta  - [ {(1 - x)\tau  + x(1 + \gamma )} ] \alpha _{ST} T}}{{(1 - x)\tau  + x(1 + \delta )}} \\[6pt]
 \Delta _{BT}  = \frac{{(1 - x)\delta  - (1 + \delta )[ {(1 - x)\tau  + x(1 + \gamma )} ] \alpha _{ST} T}}{{(1 - x)\tau  + x(1 + \delta )}}, \\[6pt]
 \end{array}
\end{equation}
where $\delta  = \frac{{a_{BT}  - a_{ST} }}{{a_{ST} }}$, $\tau  = \frac{{c_{11,ST}  + 2c_{12,ST} }}{{c_{11,BT}  + 2c_{12,BT} }}$, $\gamma  = \frac{{\alpha _{BT}  - \alpha _{ST} }}{{\alpha _{ST} }}$ , and $\alpha _{ST}$, $\alpha _{BT}$ are the thermal expansion coefficients of BT and ST respectively.

At room temperature, $\delta=0.026$ \cite{c39} and $\delta=0.024$ \cite{c40}. Below we use  $\delta = 0.026$. The thermal expansion coefficient determines the $\delta$ temperature dependence. However, in a temperature range of 0 - 500 K the change of $\delta$ is not higher than several percents, so we will neglect this dependence and consider the $\delta$ coefficient as independent of temperature.

Note that Eqs. (7) give concentration dependence of $a$ which differ from the Vegard rule $(1-x)a_{ST} + xa_{BT}$  \cite{c39,c40}. However, the ratio of elastic moduli in our case is close to 1 ($\tau = 1.13$), so at fixed temperature deviations from the simple linear dependence are rather small. Here we may note that it is quite evident physically that lattice parameter of solid solution should depend on the relative elasticity of its constituents, so appearance of $\tau$ in the Eq. (7) seems to be quite reasonable.

The potential, Eq.(6) is the Helmholtz free energy for BST with the end-member ST and BT potentials with the order parameters which are given now for the BST lattice. The Gibbs potentials $\Phi_i = F_i - \sigma u$ (i = ST, BT, $\sigma$ is the stress) may be written by substitution of strain $u$ for the quantities determined from the equation $\sigma  = \frac{{\partial F_i }}{{\partial u}}$. From the form of Eqs. (4) and (5), it follows that only constants of the second and fourth order will be renormalized. Thus, the quadratic part of the Gibbs potential of BST will be

\begin{widetext}
\begin{equation}
\begin{array}{l}
 {\rm   }\left\{ {(1 - x)\beta _{_{1,ST} }^\sigma   + x\beta _{_{1,BT} }^\sigma   + \left( {b_{11,BT}  + 2b_{12,BT}  - b_{11,ST}  - 2b_{12,ST} } \right)\frac{{x(1 - x)\left[ {(\gamma  - \delta )\alpha _{ST} T + \delta } \right]}}{{(1 - x)\tau  + x(1 + \delta )}}} \right\}\varphi ^2  \\
  + \left\{ {(1 - x)\alpha _{_{1,ST} }^\sigma   + x\alpha _{_{1,BT} }^\sigma   + \left( {g_{11,BT}  + 2g_{12,BT}  - g_{11,ST}  - 2g_{12,ST} } \right)\frac{{x(1 - x)\left[ {(\gamma  - \delta )\alpha _{ST} T + \delta } \right]}}{{(1 - x)\tau  + x(1 + \delta )}}} \right\}p^2 , \\
 \end{array}
\end{equation}
\end{widetext}

Below we will use the Gibbs potentials $\Phi_i$.

\section{ PHASE DIAGRAM OF THE $\bm{\mathbf{Ba_{x}Sr_{1-x}TiO_3}}$  SYSTEM }

To study the solid solutions in the framework of the formalism developed above, it is necessary to determine the potential coefficients at the boundary points $x = 0$ and $x = 1$ for the $\Phi_{ST}$ , and $\Phi_{BT}$  potentials, respectively.

As a base for $\Phi_{ST}$, we take the potential up to the fourth order presented in Ref. \cite{c20} and modified in Ref. \cite{c22}. Taking into account the quantum properties of $\Phi_{ST}$, which are important at low temperatures, we assume that constant $\beta_{1,ST}^\sigma$ in the $\varphi^2$ term from Eq.(8) follows the equation \cite{c44}

\begin{equation}
\beta _{_{1,ST} }^\sigma   = B\left[ {\coth \left( {\frac{{60.75}}{T}} \right) - \coth \left( {\frac{{60.75}}{{T_a }}} \right)} \right](J/m^5)
\end{equation}
with $T_a$ = 106 K .

With experimental data from \cite{c20} $\beta _{_{11,ST} }^\sigma   = 1.69 \times 10^{50} J/m^7 $  and $\varphi = 0.69 \times 10^{-11} m $ at T = 2 K, one obtains $B = 1.7 \times 10^{28} J/m^5$. Note that this  value of $\varphi$ is associated with the angle of TiO$_6$ rotation which is equal to $2^ \circ $.

For the $\Phi_{BT}$ potential of BaTiO$_3$ we take the constants from Refs. \cite{c7,c45}. The potential with these constants satisfactorily describes the phase transitions in BaTiO$_3$. The BT crystal is stable with respect to appearance of the $\varphi$ order parameter at the R (1/2, 1/2, 1/2) point of the BZ boundary. Therefore, we add to the potential \cite{c45} the part quadratic in $\varphi$  with the constant proportional to the square of the appropriate "soft mode" frequencies at the R point. The order parameter $\varphi$ at this point is associated with oxygen ion shifts (oxygen octahedral rotation) only \cite{c46}, and Ba ions do not take part in these shifts. Thus, the ratio of this frequency squared to the constant in the $\varphi^2$ term should be the same as that in the ST crystal at the same temperature. Taking for the appropriate frequencies at T = 300 K the values \cite{c14,c47} 6 meV and 16 meV for ST and BT respectively, we obtain $\beta _{_{1,BT} }^\sigma   = {\rm 3}{\rm .7} \times {\rm 10}^{{\rm 29}} $ J/m$^5$.

We neglect the other $\varphi$-dependent terms in $\Phi_{BT}$ as compared to the $\beta _{_{1,BT} }^\sigma  \varphi ^2 $ term since there is no transition with the $\varphi$ order parameter in BT.

The coefficient in the $p^2$ term we take from Refs. \cite{c22,c24}, where this value was found from the temperature dependence of ferroelectric soft mode with account of specific behavior of SrTiO$_3$ at low temperatures:

\begin{equation}
\alpha _{_{1,ST} }^\sigma   = A\left[ {\coth \left( {\frac{{54}}
{T}} \right) - \coth \left( {\frac{{54}}
{{30}}} \right)} \right]Jm/C^2
\end{equation}

According to Ref. \cite{c22}, $A=4.05 \times 10^7 $ Jm/C$^2$. This value gives the Pm3m $\leftrightarrow$ P4mm transition line in the $T-x$ phase diagram which does not agree with experiment (almost straight line in the experiment). If the thermal hysteresis is neglected, this transition line should be determined by the line of zero coefficient in the $p^2$ term in Eq.(8).  We assume that at high temperatures $\frac{{\partial \alpha _{_{1,ST} }^\sigma  }}{{\partial T}} = \frac{{\partial \alpha _{_{1,BT} }^\sigma  }}{{\partial T}}$. In such a case $\coth (a/T) \approx T/a$, and the  $xT$  term in the first two terms at $p^2$ in Eq.(8) may be neglected, and the transition line becomes close to the straight line. Then, one obtains $A=1.8\times 10^7$ Jm/C$^2$. However the best agreement with the experiment is reached with $A=1.5\times 10^7$ Jm/C$^2$. Just this last value we will use below.

Constants from Eqs. (9), (10), along with the constants $\beta _{_{11,ST} }^\sigma $, $t _{_{11,ST} }^\sigma $ and $t _{_{12,ST} }^\sigma $ determine the behavior of ferroelectric modes at temperatures below the transition point T$_a$. Since the $\alpha _{_{1,ST} }^\sigma $ constant was changed, it is necessary to redetermine the $t _{_{11,ST} }^\sigma $, $t _{_{12,ST} }^\sigma $ constants.

To do this we used the following procedure, similar to that of Ref. \cite{c24}. Neglecting a small hard modes contributon to dielectric constant we have \cite{c3}
 \[
 \frac{1} {\varepsilon } = \frac{{\partial ^2 \Phi _{ST} }} {{\partial p^2 }} = K\omega ^2,
 \]
K being some constant scale factor, which we determine from this relation at the temperatures above the ST transition temperature T$_a$. Below T$_a$ analogous relations (with the same K) for the inverse longitudinal and transversal dielectric constants define correspondingly the temperature dependence of the frequencies of the soft A$_{1u}$ and E$_u$ modes resulting from the splitting of the original F$_{1u}$ mode in ST tetragonal phase The formulae for these split frequencies are given in Ref. \cite{c20,c24} and they depend on coupling coefficients $\text{t}_{_{11,ST} }^\sigma$ ,$\text{ t}_{_{12,ST} }^\sigma $.

\begin{figure}
\includegraphics{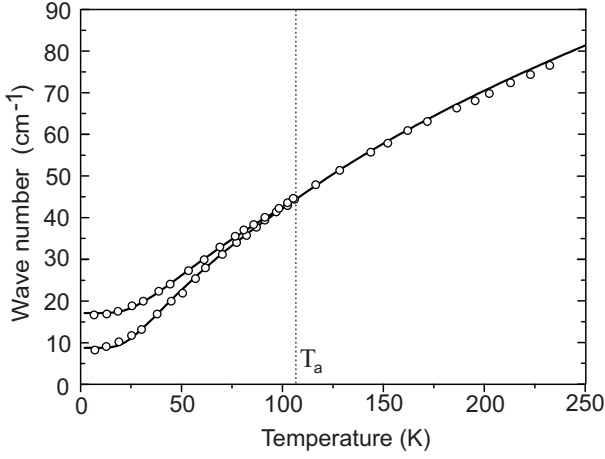}
\caption{\label{fig1} Temperature dependence of the ferroelectric modes in SrTiO$_3$. Experimental points are from Ref. \cite{c24}, solid lines are fitting with $t_{_{11,ST} }^\sigma = - 0.7 \times 10^{29} $, $t_{_{12,ST} }^\sigma = - 0.32 \times 10^{29} $ in $J/C^2m$ units and with scale value $K=2\alpha_{_{1,ST} }^\sigma/\omega^2=1.65\times 10^{4} Jm/C^2cm^{-2}$ at T=200 K.}
\end{figure}

\begin{figure*}
\includegraphics{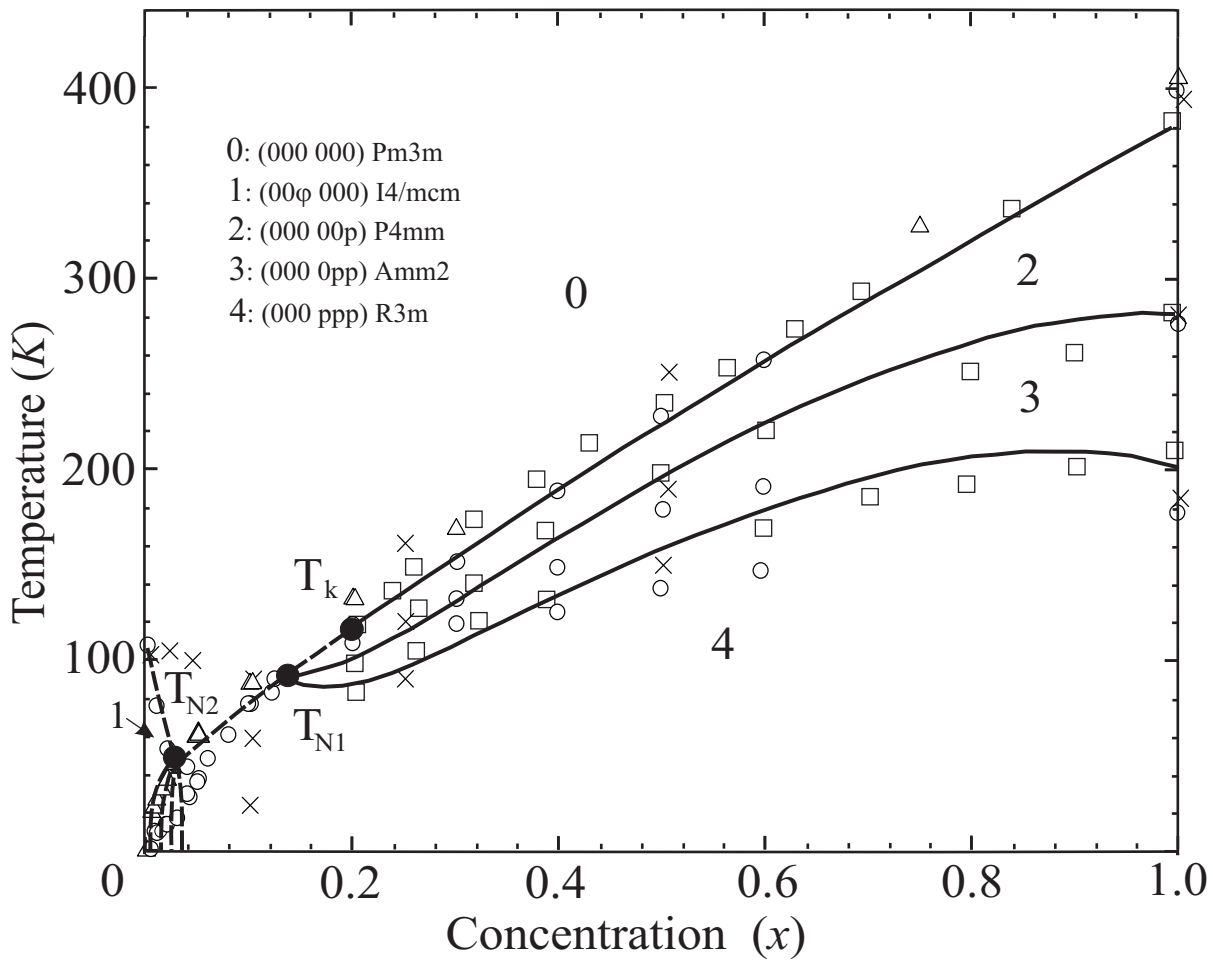}
\caption{\label{fig2}
Phase diagram of Ba$_{x}$Sr$_{1-x}$TiO$_{3}$ Solid and dashed lines present calculated diagram for the first- and second-order phase transitions, respectively. Experimental points: circles \cite{c39, c42}, crosses \cite{c41}, triangles \cite{c40}, squares \cite{c43}. The calculated coordinates of the tricritical and multiphase points are T$_k$=118 K, $x=0.2$; T$_{N1}$=93 K, $x=0.13$; T$_{N2}$=48 K, $x=0.027$. Detailed pictures of the phase diagram in the vicinity of the multiphase point T$_{N2}$ are presented in Fig. \ref{fig3}.}
\end{figure*}

Using experimental data \cite{c24}, we recalculated them to obtain  $\text{t}_{_{11,ST} }^\sigma  \text{ = } - \text{0}\text{.7} \times \text{10}^{\text{29}} $ and  $\text{t}_{_{12,ST} }^\sigma   =  - \text{0}\text{.32} \times \text{10}^{\text{29}} $ in $J/C^2m$. To examine the validity of these values, the experimental data \cite{c24} on the temperature dependence of the ferroelectric modes in SrTiO$_3$ and the dependence predicted using our values of the constants are presented in Fig.\ref{fig1} for the scale factor $K=1.65 \times 10^4 Jm/C^2cm^{-2}$.

\begin{table}
\caption{\label{tab1}
Constants of the potentials for the ST and BT crystals. Constants with superscript u are for the Helmholtz free energy $F_i=F_i(\varphi, p, u)$, constants with superscript $\sigma$, for the Gibbs potential $\Phi_i=\Phi_i(\varphi, p, \sigma)$, (i = ST, BT, u is the strain, $\sigma$, the stress). Constants in the quadratic terms: $\beta_{_{1,ST}}^\sigma = 1.7 \times 10^{28} [coth(60.75/T) - coth(60.75/106)] $, $\beta _{_{1,BT}}^\sigma = 3.7 \times 10^{29}$ in $J/m^5$ units, $\alpha _{_{1,ST}}^\sigma  = 1.5 \times 10^7 [coth(54/T) - coth(54/30)]$,  $\alpha _{_{1,BT}}^\sigma  = 3.34 \times 10^5 (T - 381)$ in $Jm/C^2$ units.
}
\begin{ruledtabular}
\begin{tabular}{cccc}
Constant& SrTiO$_3$ & BaTiO$_3$ & Units  \\ \hline \\
$\beta_{_{11}}^u$&$1.94^a$&0& $\times 10^{50} J/m^7$\\
$\beta_{_{12}}^u $&$3.97^a$&0& \\
$\beta_{_{11}}^\sigma$&$1.69^a$&0& \\
$\beta_{_{12}}^\sigma$&$3.88^a$&0& \\
$\alpha_{_{11}}^u$&0.63 ($2.0^a$)&-1.48+0.00469T$^b$& $\times 10^{9} Jm^5/C^4$\\
$\alpha_{_{12}}^u$&0.02 ($1.19^a$)&-0.0877$^b$& \\
$\alpha_{_{11}}^\sigma$&0.33 ($1.7^a$)&-2.04+0.00469T$^b$& \\
$\alpha_{_{12}}^\sigma$&0.2 ($1.37^a$)&0.323$^b$& \\
$\alpha_{_{111}}^u, \alpha_{_{111}}^\sigma $&0&24.45-0.0552T$^b$& $\times 10^{9} Jm^9/C^6$\\
$\alpha_{_{112}}^u, \alpha_{_{112}}^\sigma $&0&4.47$^b$& \\
$\alpha_{_{123}}^u, \alpha_{_{123}}^\sigma $&0&4.91$^b$& \\
$t_{_{11}}^u$&-2.0 ($-3.0^a$)&0& $\times 10^{29} J/C^2m$\\
$t_{_{12}}^u$&0.67 ($0.23^a$)&0& \\
$t_{_{11}}^\sigma$&-0.7 ($-1.74^a$)&0& \\
$t_{_{12}}^\sigma$&-0.32 ($-0.75^a$)&0& \\
$c_{11}$&3.36$^a$&2.06$^c$& $\times 10^{11} J/m^3$\\
$c_{12}$&1.07$^a$&1.4$^c$& \\
$c_{44}$&1.27$^a$&1.26$^c$&\\
$b_{11}$&1.25$^a$&0& $\times 10^{30} J/m^5$\\
$b_{12}$&-2.5$^a$&0&\\
$b_{44}$&-2.3$^a$&0& \\
$g_{11}$&1.25$^a$&1.0$^b$& $\times 10^{10} Jm/C^2$\\
$g_{12}$&-0.108$^a$&-0.017$^b$& \\
$g_{44}$&0.243$^a$&0.365$^b$& \\
\end{tabular}
\end{ruledtabular}

\footnotetext{
$^a$ Ref. \cite{c22}, $^b$ Ref. \cite{c45}, $^c$ Ref. \cite{c52}.
}
\end{table}

We discuss further our values of the constants $\alpha _{_{11,ST} }^\sigma $, and $\alpha _{_{12,ST} }^\sigma $. In the experimental phase diagram \cite{c39,c42} one observes the convergence of the lines of ferroelectrics phase transitions and probably disappearance of the P4mm and Amm2 phases. This can be described by the sign change of the anisotropic coefficient of the fourth-order invariant in $\Phi_{BST}$ which is equal to ($\alpha _{_{12} }^\sigma   - 2\alpha _{_{11} }^\sigma  $). Since $\alpha _{_{12,BT} }^\sigma   - 2\alpha _{_{11,BT} }^\sigma   > 0$, the value of ($\alpha _{_{12} }^\sigma   - 2\alpha _{_{11} }^\sigma  $) in $\Phi_{BST}$  potential should change the sign as a function of concentration $x$ (this follows from the global stability of the BT rhombohedral phase ), which  is possible if $\alpha _{_{12,ST} }^\sigma   - 2\alpha _{_{11,ST} }^\sigma   < 0$ in $\Phi_{ST}$. Moreover, one observes disappearance of thermal hysteresis for the Pm3m $\leftrightarrow$ P4mm phase transition at $x = 0.2\div 0.4$ \cite{c39,c43,c48}, i.e., there is a tricritical point in $(x-T)$ phase diagram.

We assume that three phase transition lines converge at $x = 0.13$ and  the tricritical point is observed at $x = 0.2$ ($x=0.2\div0.25$ in \cite{c39}). As a result one obtains $\alpha _{_{11,ST} }^\sigma   = 0.33$ and $\alpha _{_{12,ST} }^\sigma   = 0.2$  in $10^9Jm^5/C^4$. (If the tricritical point occurs at $x = 0.4$ \cite{c43,c48}, the constants increase up to $\alpha _{_{11,ST} }^\sigma   = 0.69$ and $\alpha _{_{12,ST} }^\sigma   = 0.92$ in $10^9Jm^5/C^4$).

All the constants in the $\Phi_{ST}$ and $\Phi_{BT}$ potentials accepted in this work are presented in Table \ref{tab1}.

For the constants presented in Table \ref{tab1}, the coordinates of the convergence point are given as $x = 0.13$, T = 93 K, and these of the tricritical point are $x = 0.2$, T = 118 K. In the calculated phase diagram there appears one more N-phase point with the coordinates $x = 0.027$, T = 48 K. The theoretical phase diagram together with experimental points \cite{c39,c40,c41,c42,c43} is shown in Fig.\ref{fig2}. The calculations both analytical and numerical were performed with the Maple-6 software package.

Symmetry of the phases and their positions near the second  N-phase point  depend on the constant $t_{44}^\sigma  $  which seems to be not quite well determined in the experiment contrary to $t_{11}^\sigma $  and $t_{12}^\sigma  $. Depending on the sign and value of $t_{44}^\sigma  $, the following three phases can appear near the N-phase point with temperature decreasing: $\left( \varphi ,\varphi ,\varphi ,{\rm  }p,p,p \right)$ - R3c, $\left( 0 ,0 ,\varphi ,{\rm  } p,p,0 \right)$ - Ima2, and $\left( \varphi _1 ,\varphi _1 ,\varphi _2 ,{\rm  p}_1 ,p_1 ,p_2 \right)$ - Cc.

\begin{figure}
\includegraphics{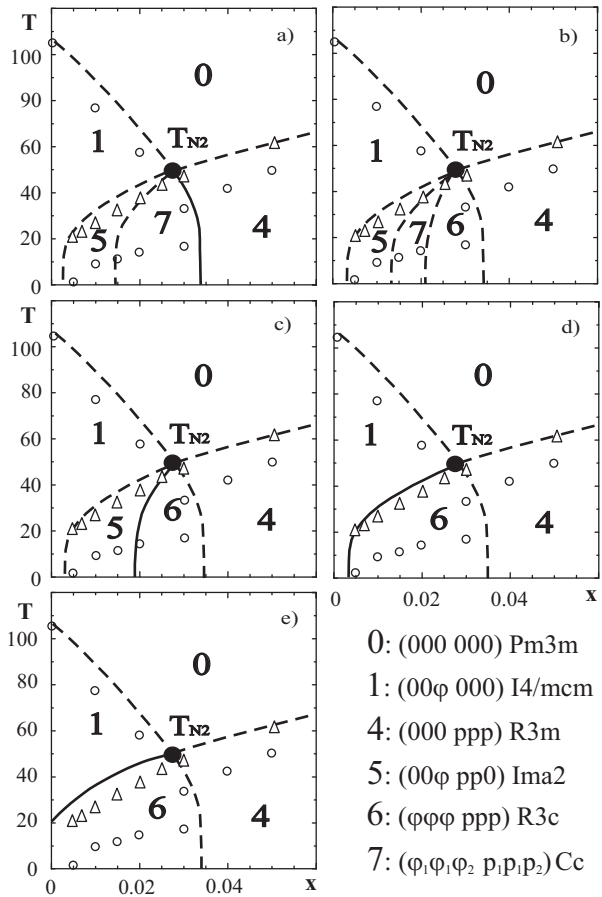}
\caption{\label{fig3} Phase diagram of Ba$_{x}$Sr$_{1-x}$TiO$_{3}$ at low $x$ in the vicinity of the $T_{N2}=48$ K multiphase point. As in Fig. \ref{fig2}, solid and dashed lines present calculated diagrams for the first- and second-order phase transitions, respectively.Experimental points are from Refs.\cite{c39,c42} (circles) and from Ref.\cite{c40} (triangles). The values of $t_{_{44}}^\sigma$ in figures a)-e) are: a) $t_{_{44}}^\sigma<0$, b) $0 < t_{_{44}}^\sigma  < 2.3 $, c) $ 2.3 < t_{_{44}}^\sigma < 2.8 $, d) $ 2.8 < t_{_{44}}^\sigma < 4 $, e) $ t_{_{44}}^\sigma > 4$ (in $\times 10^{29} J/C^2m$ units). }
\end{figure}

With negative $t_{44}^\sigma$, only the phases Ima2 and Cc exist (Fig.\ref{fig3}a), and transitions to the Cc phase are only of first order. At small positive $t_{44}^\sigma$, the phase R3c appears in a narrow region between the phases Cc and R3m (Fig.\ref{fig3}b). The phase borders between R3m $\leftrightarrow$  R3c and R3c $\leftrightarrow$  Cc phases are second-order. As $t_{44}^\sigma$ increases, the region of existence of the R3c phase becomes broader because of decreasing of the existence region of the low-symmetry phase Cc unless the latter disappears completely. In this case, the transition line between the R3c and Ima2 phases is of first order (Fig. \ref{fig3}c). As $t_{44}^\sigma$ increases further, the Ima2 phase becomes "swallowed up" by the R3c phase, and first-order phase transition is possible between the I4/mcm and R3c phases (Fig. \ref{fig3}d). At  $t_{44}^\sigma>3.1\times 10^{29}$ J/C$^2$m, the $\Phi_{ST}$ $(x = 0)$ potential describes the R3c phase at low temperatures, which at first is metastable and then at $t_{44}^\sigma>4.0\times 10^{29}$ J/C$^2$m becomes stable.

\section{DISCUSSION }
Now, we discuss the phase diagrams in Figs. \ref{fig2}, \ref{fig3}. One can see that in the region of small Ba concentration $x$, there is rather a large scatter of the experimental points. The scatter seems to be due to different methods of sample synthesis. E.g., changing temperature of the synthesis, one obtains different dielectric properties in the samples with the same concentration $x$ \cite{c49} which should lead to different values of the constant of the potentials. In Fig. \ref{fig2}, the experimental point at $x = 0.1$ from Ref. \cite{c41}  cannot be used, otherwise $\alpha _{11,ST} $ should be negative. In this case more complete model is needed with up to the sixth order in $\Phi_{ST}$.

In the phase diagrams in Figs. \ref{fig2}, \ref{fig3}, the existence of N-phase points T$_{N1}$ and T$_{N2}$ do not contradict the Gibbs phase rule. This follows from the developing of the phenomenological theory itself: along the lines and in  the points  of  second-order phase transitions, a system is in a single-phase state, namely in the phase with the highest symmetry. One should apply the Gibbs rule only to the solid lines in Figs. \ref{fig2}, \ref{fig3} (first-order transitions); in the points  of these lines, the system is multi-phase, in our case, two-phase.

Various versions of the phase diagram at small $x$ are presented in Fig. \ref{fig3}. The steepness of decreasing the $\bm{0} \leftrightarrow \bm{1}$ (Pm3m $\leftrightarrow$ I4/mcm) phase transition line as a function of $x$ is determined by the frequency of the $\varphi$ - normal mode at the R (1/2, 1/2, 1/2) point from the BZ boundary which in the BT crystal is high and is assumed to be temperature-independent.

For the line $\beta _{1}= 0 $ (the $\bm{0} \leftrightarrow \bm{1}$ phase transition line), one obtains $x= 0.034$ at $T= 0$. If the position of the $\bm{0} \leftrightarrow \bm{2}$ (Pm3m  $\leftrightarrow$  P4mm) phase transition line (see Fig. \ref{fig2}) changes depending on the sample synthesis, the N2-phase point would change its coordinates along the line $\beta _{1} = 0$. If T$_{N2}$ decreases, the region of stability of the phase $\bm{1}$ (I4/mcm) will increase up to $x = 0.034$. If T$_{N2}$ increases, there is a possibility of crossing the borderline of the phases $\bm{5}$, $\bm{6}$, $\bm{7}$ and the line $x = 0$. Thus, several phase transitions in the pure ST crystals become possible. Impurities change the deformation potential \cite{c50}  and that results in renormalization of the constants of the complete potential and in the shift of points in the phase diagram.

Let us pay attention to the phase borderlines in Figs. \ref{fig2}, \ref{fig3} which are close to vertical. Even with a very precise determination of the Ba concentration, there is always concentration ingomogeneity. That is why the vertical borderlines will be smeared. We would like also to note that another disturbance of the phase diagram may be due to possible spinodal decomposition of the solid solution and formation of BT nanoregions in the process of sample preparation \cite{c51}. Note that maybe just such a decomposition of the solid solution was the reason of glasslike behavior in the ceramic samples at low $x$ observed in Ref. \cite{c39}, though this result was not supported in the other works \cite{c37,c40,c42}.

Now, we discuss how important is the elastic contribution for the phase diagram. It appears that at, for example, $x = 0.51$, the difference between the $\bm{0}$ - $\bm{2}$ phase transition temperatures calculated with and without the elastic contribution is 24 K. At $x = 0.8$, the same value  for  the  $\bm{3}$ - $\bm{4}$ phase transition temperatures is 17 K. ( The elastic contribution decreases the phase transition temperatures.) Thus, the difference is no more than $10 \%$, i.e., not very high. However, even small variation of the potential constants can significantly change the phase diagram at small x, what is clearly seen in Fig.\ref{fig3}.

\section{ SUMMARY }
The method of construction of the potential in the framework of the phenomenological theory has been developed for description of the concentration phase diagram of non-ordering solid solutions. ( The ordered solid solution is a special case which should be treated separately.)

The concentration $(x - T)$ phase diagram of Ba$_{x}$Sr$_{1-x}$TiO$_{3}$ solid solution has been derived, and the diagram adequately agrees with the available experimental data.

There are three specific points in the phase diagram: two N-phase points and one tricritical point. The coordinates of these points critically depend on the numerical values of the potential constants, which may explain large scattering of experimental data.

At low Ba concentration, there are several phase states, which are very close with respect to the concentrations. Concentration fluctuations in the samples can result in broadening of these "quasi-vertical" borderlines.

\begin{acknowledgments}
We are grateful to P. N. Timonin for illuminating discussion during the course of this work. This study was partially supported by the Russian Foundation for Basic Research (project no. 04-02-16228) and the grant UR (01.01.270).
\end{acknowledgments}


\end{document}